\documentclass[]{pasj01}
\usepackage{xcolor}
\usepackage{amsmath}
\usepackage{multirow}
\usepackage{makecell}

\Received{}
\Accepted{}
 
\begin{document} 
\title{{Low abundances of TiO and VO on the Dayside of KELT-9 b: Insights from Ground-Based Photometric Observations}}
\author{Yuya \textsc{Hayashi}\altaffilmark{1}}

\author{Norio \textsc{Narita},\altaffilmark{2, 3, 4}}

\author{Akihiko \textsc{Fukui}\altaffilmark{2, 4}}

\author{Quentin \textsc{Changeat}\altaffilmark{5, 6}}

\author{Kiyoe \textsc{Kawauchi}\altaffilmark{7}}

\author{Kai \textsc{Ikuta}\altaffilmark{1}}

\author{Enric \textsc{Palle}\altaffilmark{4, 8}}

\author{Felipe \textsc{Murgas}\altaffilmark{4, 8}}

\author{Hannu \textsc{Parviainen}\altaffilmark{4, 8}}

\author{Emma \textsc{Esparza-Borges}\altaffilmark{4, 8}}

\author{Alberto \textsc{Pel\'aez-Torres}\altaffilmark{4, 8}}

\author{Pedro Pablo \textsc{Meni Gallardo}\altaffilmark{4}}

\author{Giuseppe \textsc{Morello}\altaffilmark{9}}

\author{Gareb \textsc{Fern\'andez-Rodr\'iguez}\altaffilmark{4, 8}}

\author{N\'estor \textsc{Abreu Garc\'ia}\altaffilmark{4, 8}}

\author{Sara \textsc{Mu\~{n}oz Torres}\altaffilmark{4, 8}}
 
\author{Y\'essica \textsc{Calatayud Borr\'as}\altaffilmark{4, 8}}

\author{Pilar \textsc{Monta\~{n}\'es Rodr\'iguez}\altaffilmark{4, 8}} 

\author{John \textsc{H. Livingston}\altaffilmark{3, 10, 11}}

\author{Noriharu \textsc{Watanabe}\altaffilmark{1}}

\author{Jerome \textsc{P. de Leon}\altaffilmark{1}}

\author{Yugo \textsc{Kawai}\altaffilmark{1}}

\author{Keisuke \textsc{Isogai}\altaffilmark{1, 12}}

\author{Mayuko \textsc{Mori}\altaffilmark{3}}

\altaffiltext{1}{Department of Multi-Disciplinary Sciences, Graduate School of Arts and Sciences, The University of Tokyo, 3-8-1 Komaba, Meguro, Tokyo 153-8902, Japan}

\altaffiltext{2}{Komaba Institute for Science, The University of Tokyo, 3-8-1 Komaba, Meguro, Tokyo 153-8902, Japan}

\altaffiltext{3}{Astrobiology Center, 2-21-1 Osawa, Mitaka, Tokyo 181-8588, Japan}

\altaffiltext{4}{Instituto de Astrof\'isica de Canarias (IAC), E-38200 La Laguna, Tenerife, Spain}

\altaffiltext{5}{Space Telescope Science Institute (STScI), 3700 San Martin Dr, Baltimore MD 21218, USA}

\altaffiltext{6}{Department of Physics and Astronomy, UCL, Gower St, Bloomsbury, London WC1E 6BT, UK}

\altaffiltext{7}{Department of Physical Sciences, Ritsumeikan University, Kusatsu, Shiga 525-8577, Japan}

\altaffiltext{8}{Departamento de Astrof\'isica, Universidad de La Laguna (ULL), E-38206 La Laguna, Tenerife, Spain}

\altaffiltext{9}{Instituto de Astrof\'isica de Andaluc\'ia (IAA-CSIC), Glorieta de la Astronom\'ia s/n, 18008 Granada, Spain}

\altaffiltext{10}{National Astronomical Observatory of Japan, 2-21-1 Osawa, Mitaka, Tokyo 181-8588, Japan}

\altaffiltext{11}{Astronomical Science Program, Graduate University for Advanced Studies, SOKENDAI, 2-21-1, Osawa, Mitaka, Tokyo, 181-8588, Japan}

\altaffiltext{12}{Okayama Observatory, Kyoto University, 3037-5 Honjo, Kamogatacho, Asakuchi, Okayama 719-0232, Japan}

\email{yuya-hayashi@g.ecc.u-tokyo.ac.jp}



\maketitle

\begin{abstract}
    We present ground-based photometric observations of secondary eclipses of the hottest known planet KELT-9b using MuSCAT2 and Sinistro. \textcolor{black}{We detect secondary eclipse signals in $i$ and $z_{\rm s}$ with eclipse depths of $373^{+74}_{-75}$ ppm and $638^{+199}_{-178}$, respectively.} We perform an atmospheric retrieval on the emission spectrum combined with the data from HST/WFC3, Spitzer, TESS, and CHEOPS to obtain the temperature profile and chemical abundances, \textcolor{black}{including TiO and VO, which have been thought to produce temperature inversion structures in the dayside of ultra-hot Jupiters}. While we confirm a strong temperature inversion structure, we find low abundances of TiO and VO with mixing ratios of \textcolor{black}{$\rm{log(TiO)}=-7.80^{+0.15}_{-0.30}$ and $\rm{log(VO)}=-9.60^{+0.64}_{-0.57}$, respectively.} The low abundances of TiO and VO are consistent with theoretical predictions for such an ultra-hot atmosphere. In such low abundances, TiO and VO have little effect on the temperature structure of the atmosphere. \textcolor{black}{The abundance of ${\rm e}^{-}$, which serves as a proxy for ${\rm H}^{-}$ ions in this study,} is found to be high, with $\rm{log(e^-)}=-4.89\pm{0.06}$. \textcolor{black}{These results indicate that the temperature inversion in KELT-9 b’s dayside atmosphere is likely not caused by TiO/VO, but rather by the significant abundance of ${\rm H}^{-}$ ions.} The best-fit model cannot fully explain the observed spectrum, and chemical species not included in the retrieval may introduce modeling biases. \textcolor{black}{Future observations with broader wavelength coverage and higher spectral resolution are expected to provide more accurate diagnostics on the presence and abundances of TiO/VO. These advanced observations will overcome the limitations of current data from HST and photometric facilities, which are constrained by narrow wavelength coverage and instrumental systematics.}
\end{abstract}

\section{Introduction}
Ultra-hot Jupiters (UHJs) are gaseous planets with equilibrium temperatures exceeding about \textcolor{black}{2000 K (e.g., \cite{Lothringer}).} Temperature inversion structures in the dayside atmospheres of UHJs have been widely detected using the Spitzer Space Telescope, Hubble Space Telescope (HST), and recently, the James Webb Space Telescope (JWST) (e.g., \cite {Deming}; \cite {changeat_edwards2022}; \cite{wasp18b-Louis-Philippe}).\par
These temperature inversions are thought to be caused by the absorption of stellar radiation by molecules such as TiO and VO, which have large cross-sections at optical wavelengths (\cite {Hubney2003}; \cite {Fortney2008}). These molecules were detected in some of the dayside atmospheres of UHJs with temperature inversions using HST/WFC3 (e.g., \cite {wasp33b-haynes}; \cite {wasp121b-changeat}), however, there has so far been no detection in the dayside atmospheres of UHJs using ground-based high-resolution cross-correlation spectroscopy (HRCCS) except for one controversial detection reported in WASP-33\ b (\cite {stev2017}). \textcolor{black}{It is worth noting that TiO and VO have been detected in the transmission spectra of WASP-189 b (\cite{WASP-189b_TiO}) and WASP-76 b (\cite{WASP-76b_VO}), respectively, using HRCCS. These detections demonstrate the reliability of the TiO and VO line databases.}\par
\textcolor{black}{Another possible optical absorber in the dayside of UHJs is the $\rm{H}^-$ ion, which is produced by the combination of hydrogen atoms from the thermal dissociation of molecular hydrogen and free electrons released by the ionization of metal atoms (\cite{bell_and_cowan}; \cite{Lothringer}; \cite{parmentier}). The bound-free absorption by $\rm{H}^-$ ions can significantly contribute to the optical opacity in these atmospheres, potentially causing temperature inversions. The flat cross-section of $\rm{H}^-$ between 1.1 and 1.4 $\mu\rm{m}$, the only wavelength range where HST/WFC3 is sensitive to TiO, suggests that the presence of $\rm{H}^-$ ions could mask the spectral features of TiO. \textcolor{black}{These factors, along with ${\rm H_2O}$ dissociation at high temperatures, contribute to making emission spectra appear black-body-like and featureless} (e.g., \cite{Kreidberg}; \cite{Mansfield18}; \cite{Arcangeli}; \cite{Mansfield21}).}\par
\textcolor{black}{Atomic metals such as iron can also produce the temperature inversion structure by absorbing stellar radiation shorter than $0.5\ \mu \rm{m}$ (\cite{Lothringer}; \cite{Lothringer-2019}) and these \textcolor{black}{species} are widely detected in the dayside of UHJs using HRCCS (e.g., \cite{Hoeijmakers-2022}, \cite{Borsa-2022}, \cite{Herman-2022}).}

In this paper, we report the secondary eclipses observations of KELT-9 b using ground-based telescopes with MuSCAT2 and Sinistro. KELT-9 b, the hottest planet known so far, orbits an A0 star ($T_{\rm eff}=10,140$ K) on a 1.48-day orbit and its dayside brightness temperature is $\sim$ 4600 K (\cite{Gaudi}). Previous observations have detected the presence of a thermal inversion on its dayside (\cite{mansfield20}). \textcolor{black}{\citet{changeat2021} and \citet{changeat_edwards2022} suggested that there are large amounts of TiO ($\rm{log(TiO)}=-6.8^{+0.3}_{-0.3}$) and VO ($\rm{log(VO)}=-6.6^{+0.3}_{-0.2}$) based on the observation with HST/WFC3. For reference, \citet{Kitzmann} predicted that for a solar composition atmosphere at 0.1 bar (approximately the pressure level of the observed photosphere) and a temperature of 4000 K, the TiO abundance would decrease to around $\rm{log(TiO)} \approx -10 $ due to thermal dissociation.} High abundances of those molecules, which were found to be important to explain spectral modulations of the HST spectrum, are therefore surprising. \citet{Kasper} set an upper limit of $\rm{log(TiO)}<-8.5$ at 99\% confidence using HRCCS technique and \citet{Ridden-Harper} also reported the non-detection of TiO and VO.
\textcolor{black}{\citet{Jacob} applied a correction for the stellar pulsation effect, which was not included in the analysis by \citet{changeat2021} and performed an equilibrium chemistry retrieval. They derived extremely high metallicity (${\rm [M/H]}=1.98$) and low C/O ratio (C/O=0.14), but this situation is unlikely and led them to consider the possibility of a quench (disequilibrium) effect even in environments hot enough to assume that equilibrium chemistry is valid (\cite{Kitzmann}; \cite{parmentier}). By considering the quench effect, the large abundances of TiO and VO found by \citet{changeat2021} could also potentially be explained by the mixing of these molecules from deeper layers of the atmosphere, where they are stable under chemical equilibrium conditions. In their disequilibrium retrieval result, KELT-9 b has nearly solar metallicity and C/O ratio while the TiO abundance ($\rm{log(TiO)}=-7.9$) contradicts that of the upper limit derived by \citet{Kasper}.}
Moreover, even their best-fit model does not fully explain the HST/WFC3 spectrum. \textcolor{black}{In addition, the narrow wavelength coverage of HST might not be able to fully break the degeneracies between the abundances of key species (TiO, VO, FeH, H2O), the thermal structure, and thermal dissociation processes ($\rm{H}^-$). When combining with other datasets, especially photometric points such as Spitzer, instrumental systematics could render the datasets incompatible (\cite{Yip}; \cite{changeat-2020}; \cite{colon}).}\par
\textcolor{black}{This paper presents a thermal emission spectrum using ground-based telescopes in the wavelength range of $0.4$ - $1.1\ \mu \rm{m}$, which is shorter than the HST/WFC3 wavelength coverage.} Section \ref{sec:obs} provides details of the observations and reductions, followed by the light curve analyses in section \ref{sec:ana}. In Section \ref{sec:ret}, we analyze the atmospheric properties of KELT-9 b, with a summary provided in Section \ref{sec:sum}.

\section{Observation and Data Reduction}
\label{sec:obs}
\subsection{MuSCAT2 photometry}
\begin{figure}
 \begin{center}
  \includegraphics[width=8cm]{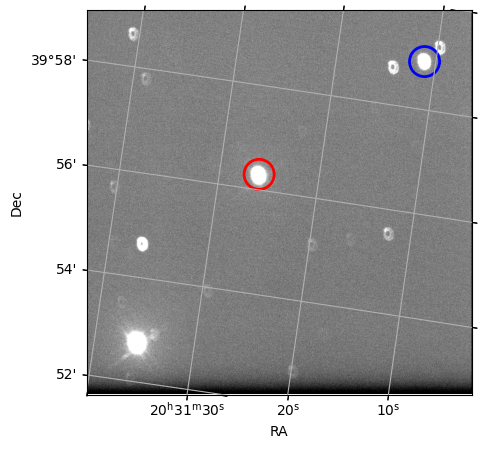} 
 \end{center}
\caption{MuSCAT2 $i$-band full-frame image during the observation. KELT-9 is marked with a red circle and the reference star is marked with a blue circle.}\label{fig:MuSCAT2FOV}
\end{figure}

We observed one secondary eclipse of KELT-9\ b with MuSCAT2 (\cite{Narita-M2-JATIS}) on the 1.52\ m TCS telescope at the Teide Observatory in the Canary Islands, Spain, on UT 2018 August 10. MuSCAT2 can take image with $g$ ($0.40$ - $0.55$ ${\mu \rm{m}}$), $r$ ($0.55$ - $0.70$ ${\mu \rm{m}}$), $i$ ($0.70$ - $0.82$ ${\mu \rm{m}}$), and $z_{\rm s}$ ($0.82$ - $0.92$ ${\mu \rm{m}}$) bands, simultaneously. Each channel has a ${\rm 1k}\times{\rm 1k}$ CCD camera, with a pixel scale of $\sim$ 0.44 arcsec, providing a field of view (FOV) of $7.4\times7.4\ {\rm arcmin^2}$. During the observation, we heavily defocused the telescope to prevent saturation. The typical value of the full width at half maximum (FWHM) of the stellar point spread function (PSF) was $\sim 13$ arcsec. The detailed information about the observation is given in table \ref{table:M2andSinistro-observation}. Figure \ref{fig:MuSCAT2FOV} is a full-frame image of $i$-band showing the observing FOV. The target star, KELT-9, and the reference star are marked as a red circle and a blue circle, respectively. \par
After dark current and flat-field calibrations, we conducted differential photometry using a custom pipeline presented in \citet{M2-reduction-pipeline}.

\subsection{Sinistro photometry}
\begin{figure}
 \begin{center}
  \includegraphics[width=8cm]{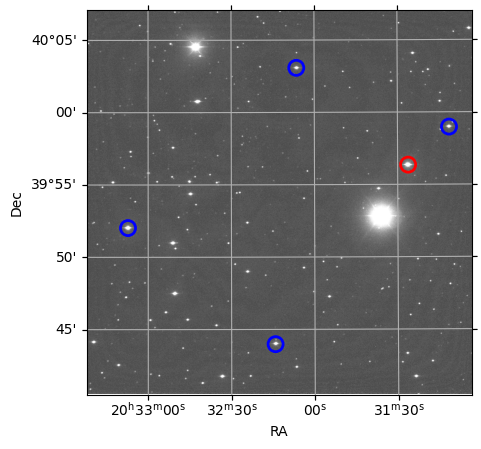} 
 \end{center}
    \caption{Sinistro $Y$-band full-frame image during the observation. KELT-9 is marked with a red circle and the reference stars are marked with blue circles.}\label{fig:SinistroFOV}
\end{figure}

We also observed two secondary eclipses with Sinistro (\cite{sinistro}) on the 1\ m Las Cumbres Observatory (LCO) telescope, located at the same site as MuSCAT2, on UT 2023 July 15 and 21. We used a $Y$-band ($0.95$ - $1.1$ ${\mu \rm{m}}$) filter for both observations. Sinistro has a ${\rm 4k}\times{\rm 4k}$ CCD camera, with a pixel scale of $0.389$\ arcsec, providing a FOV of $26.5\times26.5\ {\rm arcmin^2}$. We defocused the telescope during the observations to avoid saturation of the detector. The typical value of the FWHM of the stellar PSF was $\sim 5$ arcsec for the two observations. The detailed information about the observations is listed in table \ref{table:M2andSinistro-observation}. Figure \ref{fig:SinistroFOV} is a full-frame image of $Y$-band showing the observing FOV.
The target star, KELT-9, and the reference stars are marked as a red circle and blue circles, respectively. \textcolor{black}{The comparison stars were selected to minimize the root mean square of the difference between data points in the extracted light curve at adjacent times similar to the procedure applied by \citet{Fukui}.}\par
The data calibration was conducted by the LCOGT ${\mathsf {
BANZAI}}$ pipeline. Then, we performed differential photometry using a custom pipeline presented in \citet{M2-reduction-pipeline}.

\begin{longtable}{ccccccc}
  \caption{Observation summary}\label{table:M2andSinistro-observation}
  \hline              
  Telescope & Instrument & Start Date (UT) & Observing time (UT) & Filter & $t_{\rm exp}$ (s) & Airmass\\ 
\endfirsthead
  \hline
  Tele. & Inst. & Start Date (UT) & Observing time (UT) & Filter & $t_{\rm exp}$ (s) & Airmass\\ 
\endhead
  \hline
\endfoot
  \hline
\endlastfoot
  \hline
  TCS 1.52\ m & MuSCAT2 & 2018 Aug 09 & 22:16-5:00 & $g, r, i, z_{\rm s}$ & 2.0, 2.0, 5.0, 10.0 & 1.14-1.02-1.83\\
  LCO 1\ m & Sinistro & 2023 Jul 15 & 23:29-6:30 & $Y$ & 30 & 1.20-1.02-1.33 \\
  LCO 1\ m & Sinistro & 2023 Jul 21 & 21:40-3:50 & $Y$ & 30 & 1.56-1.02-1.12 \\
\end{longtable}

\section{Light Curve Analysis and Result}
\label{sec:ana}
\subsection{Model Settings}
To model the secondary eclipse, we use the Python package ${\mathsf {Batman}}$ (\cite{batman}) with the following parameters: eclipse depth $f_{\rm p}$, ratio of semi-major axis to stellar radius $a/R_{\rm s}$, eclipse center time $T_{\rm c}$. We assume a circular orbit. We allow $a/R_{\rm s}$ and $T_{\rm c}$ to be shared among the four ($g$, $r$, $i$, and $z_{\rm s}$) light curves of MuSCAT2. In Sinistro light curves, $a/R_{\rm s}$ and $f_{\rm p}$ are to be shared within the two observations. The orbital period $P$, the orbital inclination $i$, and the ratio of planet radius to stellar radius $R_{\rm p}/R_{\rm s}$ are fixed to literature values ($P=1.4811$ days, $i=87\degree$, $R_{\rm p}/R_{\rm s}=0.0823$; \cite{Gaudi}) in the eclipse model.\par
In addition, we also employ a Gaussian process (GP) to model time-correlated systematic noise in the light curves using ${\mathsf {
celerite}}$ (\cite{celerite}) with an approximated Mat\'ern-3/2 covariance matrix:
\begin{equation}
\begin{split}
\kappa(\tau)&=\frac{\sigma_{\rm GP}^2}{2}\left[\left(1+\frac{1}{\epsilon}\right)e^{-(1-\epsilon)\sqrt{3}\tau/\rho}+\left(1-\frac{1}{\epsilon}\right)e^{-(1+\epsilon)\sqrt{3}\tau/\rho}\right] \\
&+\sigma^2_w\delta,
\end{split}
\end{equation}

where $\tau=|t_i-t_j|$ is the distance of time between two data points, $\sigma_{\rm GP}$ and $\rho$ are the amplitude and length scales of the systematic noise, $\epsilon$ controls approximation quality, $\sigma^2_w$ denotes the white noise of each data point. We let ${\rm ln}\, \sigma_{\rm GP}$ to be free for each light curve while we set $\rm{ln}\, \rho$ to be shared within four ($g$, $r$, $i$, and $z_{\rm s}$) light curves of MuSCAT2 assuming that a timescale of the systematic noise is common for all bands. \textcolor{black}{We set the $\epsilon$ values to $0.01$, which is the default value in celerite.}\par

We then perform an MCMC ensemble sampling using the Python package ${\mathsf {emcee}}$ (\cite{emcee}) to derive the posterior probability distributions of the free parameters. We adopt uniform priors for $f_{\rm p}$ of each band, and Gaussian priors for other free parameters of the eclipse model: $T_c$ and $a/R_{\rm s}$ with values and uncertainties given by \citet{Gaudi} and \citet{Ivshina-Winn}, respectively. Note that we enlarge the uncertainties of $T_{\rm c}$ and $a/R_{\rm s}$ provided by these references by a factor of five considering the possibility of changes in the planetary orbital parameters. We apply log-uniform priors for GP hyperparameters. We set the lower bound of $\rm{ln}\, \rho$ to -2 so that the timescale of systematic noise is longer than the duration of eclipse ingress and egress. Although the host star KELT-9 has $\delta$-Scuti type stellar pulsation with a period of $\sim7.6\ {\rm hr}$ (\cite{wong-2021}), this variability is also removed by the GP along with other systematic noise.

\subsection{Result of Light Curve Analysis}
\begin{figure*}
 \begin{center}
  \includegraphics[width=16cm]{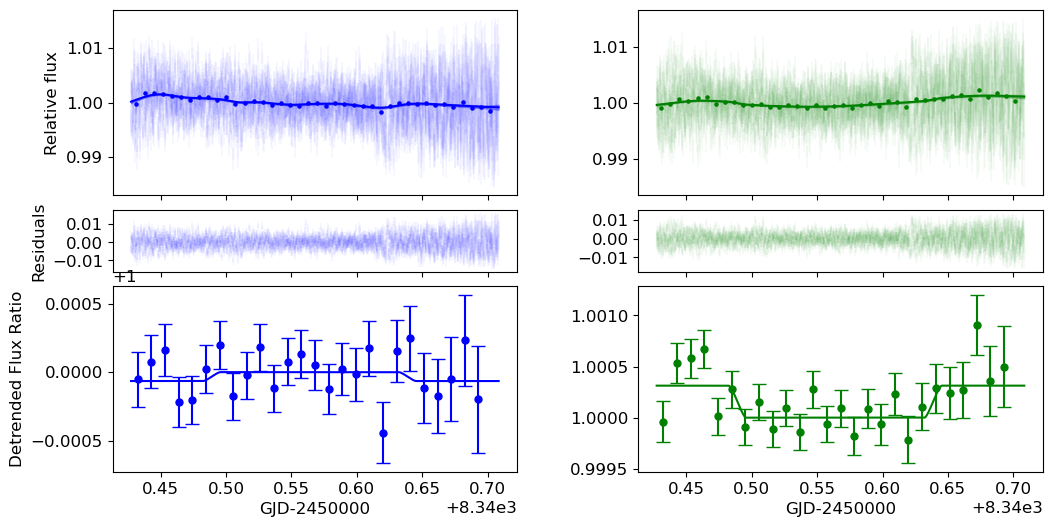}
  \includegraphics[width=16cm]{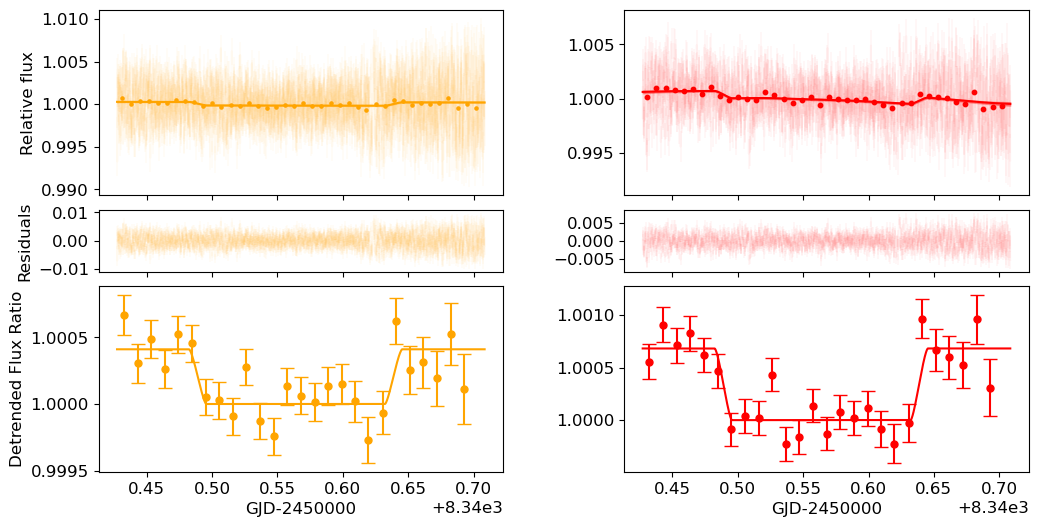} 
 \end{center}
\caption{The secondary eclipse of KELT-9 b observed by MuSCAT2 in $g$ ($0.40$ - $0.55$ ${\mu \rm{m}}$, upper-left), $r$ ($0.55$ - $0.70$ ${\mu \rm{m}}$, upper-right), $i$ ($0.70$ - $0.82$ ${\mu \rm{m}}$, lower-left), $z_{\rm s}$ ($0.82$ - $0.92$ ${\mu \rm{m}}$, lower-right) band on 2018 August 9. In each band, the top panel shows the relative flux ratio with the best-fit model. The second panel shows the residuals from the best-fit model. The bottom panel shows the detrended light curve with the best-fit GP model. Circles in the top panels and data points in the bottom panels are bins of observation data every 900 seconds.}\label{fig:MuSCAT2_light-curve}
\end{figure*}

\begin{figure*}
 \begin{center}
  \includegraphics[width=16cm]{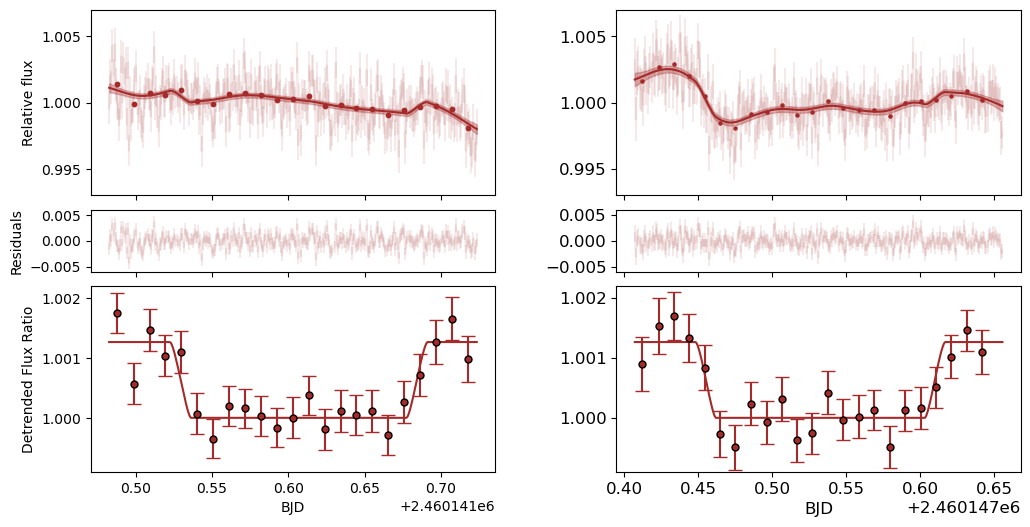} 
 \end{center}
\caption{The secondary eclipse of KELT-9 b observed by Sinistro in $Y$-band ($0.95$ - $1.1$ ${\mu \rm{m}}$) on 2023 July 15 (left) and 2023 July 21 (right). Same as figure \ref{fig:MuSCAT2_light-curve}.}\label{fig:Sinistro_light-curve}
\end{figure*}

The best-fit light curves of MuSCAT2 data are shown in figure \ref{fig:MuSCAT2_light-curve} and derived eclipse parameters are listed in table \ref{tab:eclipse_parameters_MuSCAT2}. Figure \ref{fig:Sinistro_light-curve} shows the best-fit light curves of Sinistro. The derived eclipse parameters are listed in table \ref{tab:eclipse_parameters_Sinistro}.\par
\textcolor{black}{We detect secondary eclipse in $i$ and $z_{\rm s}$-band with $5.0\sigma$ and $3.6\sigma$ significance, respectively. Although the detections in the $g$, $r$, and $Y$-bands do not reach the $3\sigma$ threshold, we include these bands in the subsequent analysis.}
Changing the kernel function to the SHO kernel does not significantly affect the results, as the derived parameters are consistent within $1\sigma$. 
We also attempt to fit the data using a prior distribution with $T_{\rm c}$ and $a/R_{\rm s}$ uncertainty 10 times larger than in previous studies The parameters obtained are consistent within $1\sigma$ for the MuSCAT2 data, even when assuming 10 times larger error bars. \textcolor{black}{However, for the Sinistro data, the fit do not converge when assuming 10 times larger error bars.} \par

\begin{table}
  \tbl{Derived secondary eclipse parameters for MuSCAT2}{%
  \begin{tabular}{ccc}
      \hline
      Parameter & Prior & Value \\ 
      \hline
      $T_{\rm c}$\footnotemark[$*$] & ${\mathcal N}(8340.5661,\, 0.0005)$ & \textcolor{black}{$8340.5654\pm{0.0005}$} \\
      $f_{\rm p}(g)$\footnotemark[$\dag$] & \multirow{4}{*}{${\mathcal U}(\textcolor{black}{-10^6},\, 10^6)$} & \textcolor{black}{$53^{+442}_{-489}$} \\
      $f_{\rm p}(r)$\footnotemark[$\dag$] &  & \textcolor{black}{$514^{+288}_{-322}$} \\
      $f_{\rm p}(i)$\footnotemark[$\dag$] & $                                   $ & \textcolor{black}{$373^{+74}_{-75}$} \\
      $f_{\rm p}(z)$\footnotemark[$\dag$] & $                                   $ & \textcolor{black}{$638^{+199}_{-178}$} \\
      $a/R_{\rm s}$ & ${\mathcal N}(3.153,\, 0.050)$ & \textcolor{black}{$3.166^{+0.043}_{-0.044}$} \\
      \hline
    \end{tabular}}\label{tab:eclipse_parameters_MuSCAT2}
    \begin{tabnote}
    \footnotemark[$*$] BJD-2450000.\\ 
    \footnotemark[$\dag$] Unit: ppm. \\
    \end{tabnote}
\end{table}

\begin{table}
  \tbl{Derived secondary eclipse parameters for Sinistro}{%
  \begin{tabular}{ccc}
      \hline
      Parameter & Prior & Value \\ 
      \hline
      $T_{\rm c1}$\footnotemark[$*$] & ${\mathcal N}(10141.606,\, 0.001)$ & \textcolor{black}{$10141.606\pm0.001$} \\
      $T_{\rm c2}$\footnotemark[$\dag$] & ${\mathcal N}(10147.530,\, 0.001)$ & \textcolor{black}{$10147.530\pm0.001$} \\
      $f_{\rm p}$\footnotemark[$\ddag$] & ${\mathcal U}(\textcolor{black}{-10^6},\, 10^6)$ & \textcolor{black}{$895^{+428}_{-411}$} \\
      $a/R_{\rm s}$ & ${\mathcal N}(3.153,\, 0.050)$ & \textcolor{black}{$3.146^{+0.047}_{-0.045}$} \\
      \hline
    \end{tabular}}\label{tab:eclipse_parameters_Sinistro}
    \begin{tabnote}
    \footnotemark[$*$] BJD-2450000, first visit (2023 Jul 15).  \\ 
    \footnotemark[$\dag$] BJD-2450000, second visit (2023 Jul 21). \\
    \footnotemark[$\ddag$] Unit: ppm. \\ 
    \end{tabnote}
\end{table}

\section{Atmospheric Retrieval}
\label{sec:ret}
There are two contributions to the eclipse depth: thermal emission and reflection. We cannot break this degeneracy, however, considering the low geometric albedo ($A_{\rm g}<0.14$) in NUV (\cite{Hooton}), the reflection light component is estimated to be $<95$ ppm. Hereafter, we assume that the derived eclipse depths are only from the thermal emission.
\subsection{Atmospheric Model}
To retrieve components in the dayside atmosphere of KELT-9 b, we attempt to model the eclipse depths derived in this work (blue) in combination with those from HST/WFC3 (\cite{Jacob}), Spitzer 3.6 $\mu \rm{m}$ (Beatty et al. in prep.), Spitzer 4.5 $\mu \rm{m}$ (\cite{mansfield20}), TESS, and CHEOPS (\cite{Jones}) using open-source retrieval code ${\mathsf {TauREx3}}$ (\cite{TauREx3}). We confirmed that the TESS apertures were not contaminated by other bodies. It is known that combining and interpreting data obtained from different instruments may not lead to correct conclusions because of instrumental systematics (e.g., \cite{Yip}). This is mainly due to the dependence of the derived white transit/eclipse depth to instrument systematic assumptions with HST. We then include an offset parameter for the HST/WFC3 data with a uniform prior, ${\mathcal U}(-300,\, 300)$ ppm. \textcolor{black}{Several studies indicated that there may be an offset of several tens of ppm between the obtained eclipse depth value in combination with the phase curve and the value obtained from the secondary eclipse alone (\cite{Jones}; \cite{Morello_2019}, \cite{Morello_2023}) due to the curvature in the planetary flux so called phase-blend (\cite{phase-blend}).} We then also include an offset parameter for the MuSCAT2 and Sinistro data with a uniform prior, ${\mathcal U}(-100,\, 100)$ ppm. We divide the planetary atmosphere from $10^7$ to $10^{-5}$ Pa into 100 layers in log space and assume hydrostatic equilibrium. We implement the Npoint profile (N=2) (\cite{2layers}) to model the temperature inversion. The temperature and pressure priors for each layer are ${\mathcal U} (1000, 8000)$ K and ${\mathcal J} (10^{-5}, 10^7)$ Pa, respectively. We retrieve chemical abundances using a chemical equilibrium model due to the high-temperature environment (\cite{Kitzmann}; \cite{parmentier}). \textcolor{black}{\citet{Jacob} and \citet{changeat_edwards2022} suggested that disequilibrium processes might play a role in the atmosphere of KELT-9 b, despite the high temperatures. To account for this possibility and to directly retrieve the abundances without assuming chemical equilibrium, we also perform a retrieval using a free chemistry model, where the abundances are assumed to be constant with pressure.} In the equilibrium chemistry framework, we use ${\mathsf {Fastchem}}$ (\cite{fastchem}) implemented by the chemical equilibrium plugins for TauREx (\cite{TauREx-Fastchem_plugins}). We set the metallicity and C/O ratio as free parameters and impose a prior of $\mathcal{U} (-2, 2)$ and $\mathcal{U} (0.1, 2)$, respectively.
In the free chemistry framework, we take into account $\rm{e^-}$ (\cite{John}), $\rm{H_2O}$ (\cite{H2O}), $\rm{CO}$ (\cite{CO}), $\rm{FeH}$ (\cite{FeH1}; \cite{FeH2}; \cite{FeH3}), $\rm{TiO}$ (\cite{TiO}), and $\rm{VO}$ (\cite{VO}) opacity. In our free chemistry retrievals, we fix the H abundance using a two-layer profile to calculate the $\rm{H^-}$ opacity without the degenerated parameters (H and $\rm{e^-}$) following \citet{changeat2021}. The volume mixing ratio of H is set to $10^{-6}$ at the surface and $0.9$ at the top. The mixing ratios for the other species are set as free parameters, with $\mathcal{J} (10^{-12}, 10^{-2})$ given as priors. The remaining atmospheric components are filled with hydrogen molecules and helium with He/${\rm H_2}$ = 0.17 following \citet{changeat2021}.
We use the MultiNest sampler (\cite{multinest}) with an evidence tolerance of 0.5 and \textcolor{black}{900} live points.

\subsection{Results and discussion}
\begin{figure*}[H]
 \begin{center}
  \includegraphics[width=16cm]{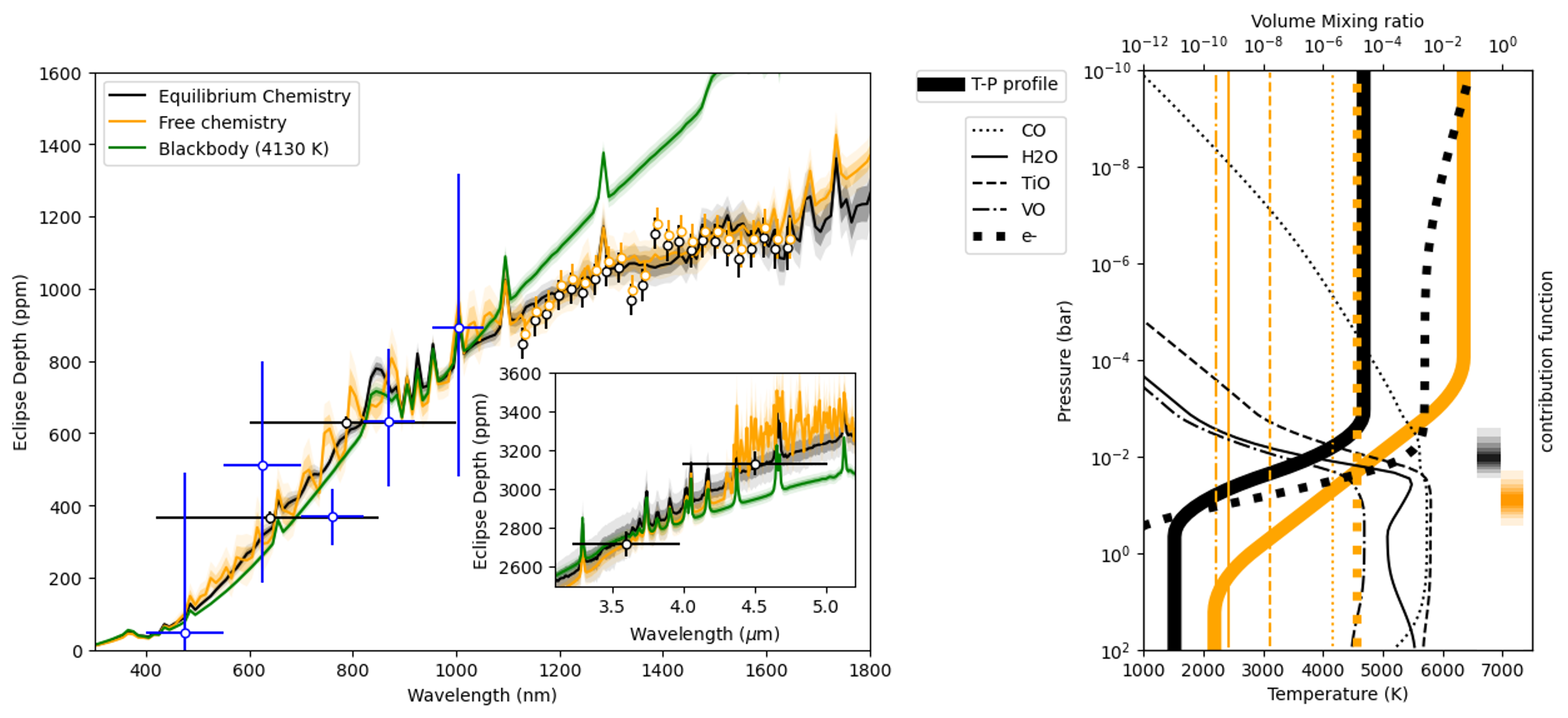}
 \end{center}
\caption{Best-fit result of atmospheric retrievals. Left: Dayside eclipse spectrum from this work (blue) with HST/WFC3 (including offsets), Spitzer, TESS, and CHEOPS (Black, except for the WFC3 data points in free chemistry, which are orange). The black, orange, and green solid lines are the model spectra of the equilibrium chemistry, free chemistry, and blackbody (4130 K), respectively. The 1$\sigma$ and 2$\sigma$ regions of each model are shown in shaded regions. Right: Retrieved temperature profiles (thick) and volume mixing ratios of CO (dotted), $\rm{H_2O}$ (solid), TiO (dashed), VO (dashed-dotted), and $\rm{e^-}$ (thick dotted), respectively. The contours on the right edge denote the contribution functions. The combination of colors is the same as the left.}\label{fig:result}
\end{figure*}

The best-fit model is shown in figure \ref{fig:result}, and the derived parameters are listed in table \ref{tab:retrieval_result}. Contributions of each species are shown in figure \ref{figA:contribution}. The posterior distribution of the free chemistry model and equilibrium chemistry model are shown in figure \ref{figA:FREE_corner} and figure \ref{figA:EQ_corner} in the Appendix, respectively. The retrieved temperature profile and chemical abundances are also shown in the same figure. \textcolor{black}{Significant temperature inversion layers ranging from 2200 K to 6400 K in the free chemistry model and from 1500 K to 4700 K in the equilibrium chemistry model are retrieved.} The equilibrium chemistry model has a higher likelihood, with a difference of \textcolor{black}{$\Delta {\rm ln}\,{\mathcal Z}=6.4$}, \textcolor{black}{compared} to the free chemistry model. We also try models with 3 and 4 points in the temperature profile. For the free chemistry model, the values of $\Delta {\rm ln}\,{\mathcal Z}$ are +0.4 and +2.7, respectively, compare to the 2-point profile. Neither of them reach the decisive criterion for model selection, $\Delta {\rm ln}\,{\mathcal Z}>4.6$, so we choose the 2-point profile based on Occam's razor. All models with different temperature profiles show temperature inversions, and the chemical abundances agree within $2\sigma$. We perform a similar comparison for the equilibrium chemistry model, and the values of $\Delta {\rm ln}\,{\mathcal Z}$ for the 3-point and 4-point profiles are +2.5 and +1.7, respectively. \textcolor{black}{Therefore, we also find no statistically significant preference for temperature-pressure profiles with more than 2 points in chemical equilibrium retrievals.} Regardless of the temperature profile used, all models exhibit temperature inversions, and the metallicity and C/O ratio values are consistent within $1\sigma$.\par

\begin{longtable}{cccccc}
\caption{Derived atmospheric parameters}\label{tab:retrieval_result}
\hline
Parameters & Free & Equilibrium & \multicolumn{3}{c}{Literature} \\ \cline{4-7}
& &  & \parbox{2cm}{\rule{0pt}{2ex}\centering Free \\ (\cite{changeat_edwards2022})} & \parbox{2cm}{\rule{0pt}{2ex} \centering Equilibrium (\cite{changeat_edwards2022})} & \parbox{2cm}{\rule{0pt}{2ex} 
\centering Equilibrium (\cite{Jacob})} \\ \hline
\endfirsthead

\hline
\endhead

\hline
\endfoot

\hline
\endlastfoot

$T_{\rm bottom}$ (K) &
\textcolor{black}{$2185^{+187}_{-250}$} &
\textcolor{black}{$1517^{+398}_{-280}$} & & & \\
$T_{\rm top}$ (K) &
\textcolor{black}{$6351^{+304}_{-325}$} &
\textcolor{black}{$4679^{+142}_{-149}$} & & &\\
${\rm log}(P_{\rm bottom})$ (Pa) &
\textcolor{black}{$5.59^{+0.19}_{-0.23}$} &
\textcolor{black}{$4.00\pm0.27$} & & &\\
${\rm log}(P_{\rm top}) $ (Pa) &
\textcolor{black}{$1.60\pm0.33$} &
\textcolor{black}{$2.72^{+0.39}_{-0.20}$} & & &\\
${\rm log(TiO)}$ & \textcolor{black}{$-7.80^{+0.15}_{-0.30}$} & & $-6.8^{+0.3}_{-0.3}$ & & \\
${\rm log(VO)}$ & \textcolor{black}{$-9.60^{+0.64}_{-0.57}$} &  & $-6.6^{+0.2}_{-0.2}$ & & \\
${\rm log(FeH)}$ & \textcolor{black}{$-8.48^{+0.30}_{-0.24}$} &  & $-8.0^{+1.1}_{-2.0}$ & & \\
${\rm log(CO)}$ & \textcolor{black}{$-5.68^{+0.64}_{-0.57}$} &  & $<-4.7$ & & \\
${\rm log(H_2O)}$ & \textcolor{black}{$-9.16^{+0.71}_{-0.87}$} &  & $<-7.0$ & & \\
${\rm log(e^-)}$ & \textcolor{black}{$-4.89^{+0.06}_{-0.06}$} &  & $-4.9^{+0.2}_{-0.1}$ & & \\
${\rm [M/H]}$ & & \textcolor{black}{$1.04^{+0.20}_{-0.34}$} & $-2.5^{+1.0}_{-1.0}$ & $1.6^{+0.2}_{-0.4}$ & $1.98^{+0.19}_{-0.22}$\\
${\rm C/O\ ratio}$ & & \textcolor{black}{$0.37^{+0.26}_{-0.19}$} & $0.6^{+0.5}_{-0.5}$ & $1.1^{+0.6}_{-0.6}$ & $0.14^{+0.10}_{-0.03}$\\
${\rm WFC3\ Offset}$ (ppm)& \textcolor{black}{$-200^{+25}_{-25}$} & \textcolor{black}{$-227^{+27}_{-21}$} & & &\\
\makecell{MuSCAT2 \& Sinistro \\ Offset (ppm) }& \textcolor{black}{$4^{+25}_{-21}$} & \textcolor{black}{$43^{+18}_{-21}$} & & &\\
${\rm ln}\,{\mathcal Z}$ & \textcolor{black}{$235.6$} & \textcolor{black}{$242.0$} & & & \\
${\rm reduced-}\chi^2$ & \textcolor{black}{$2.24$} & \textcolor{black}{$1.89$} & & & \\
\end{longtable}

\subsubsection{The effect of TiO and VO on the temperature inversion}
\begin{figure}
 \begin{center}
  \includegraphics[width=8cm]{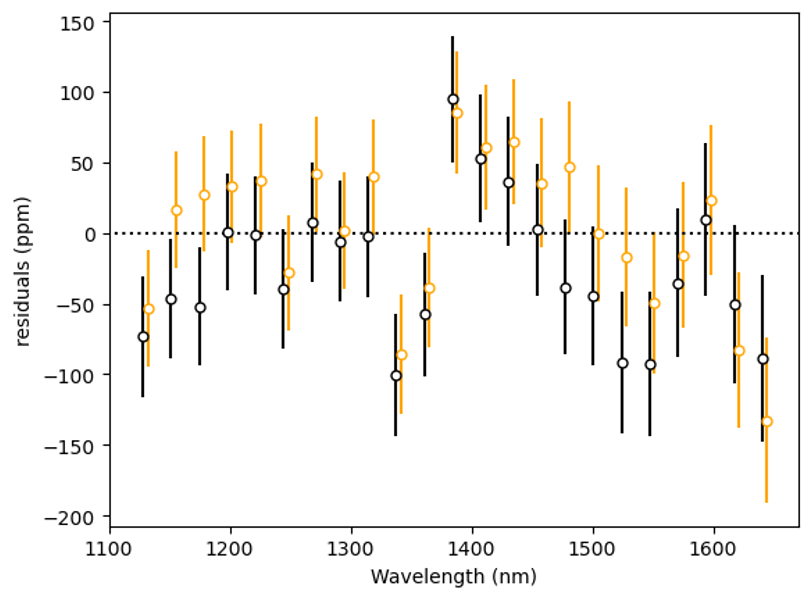} 
 \end{center}
\caption{Difference between best-fit spectrums and HST/WFC3 observed data points with offsets. The color combination is the same as in figure \ref{fig:result}.}\label{fig:residuals}
\end{figure}

In the free chemistry model, the derived abundances of TiO and VO are \textcolor{black}{$\rm{log(TiO)}=-7.80^{+0.15}_{-0.30}$ and $\rm{log(VO)}=-9.60^{+0.64}_{-0.57}$, respectively.} These values are consistent with \citet{Jacob}. \textcolor{black}{It is worth noting that the inclusion of additional data in the optical band, which includes our ground-based observations as well as the TESS and CHEOPS data points where TiO and VO have large cross-sections, does not significantly alter the derived abundances of these molecules.} Low abundances of TiO and VO are consistent with the theoretical prediction (\cite{Kitzmann}), \textcolor{black}{suggesting that these abundances are unlikely to significantly impact the temperature structure of the atmosphere (\cite{Lothringer}; \textcolor{black}{\cite{Gandhi_and_Madhusudhan_2019}}).}  \textcolor{black}{The derived TiO abundance agrees with the upper limit obtained by \citet{Kasper} using HRCCS within a $3\sigma$ range. Moreover, considering that TiO dissociates more in the higher temperature upper regions probed by HRCCS, the abundance in the upper layers would be even closer to the HRCCS results. The derived abundance of TiO is about $10^{-1}$ times lower than that reported in previous studies using the same HST/WFC3 and Spitzer data (\cite{changeat2021}; \cite{changeat_edwards2022}). Similarly, the derived abundance of VO is about $10^{-3}$ times lower than that in the same studies. This may be due to the difference between the correction of stellar pulsation which is done by \citet{Jacob}.}
In the free chemistry model, \textcolor{black}{the model without TiO/VO shows a worse fit ($\Delta {\rm ln}\,{\mathcal Z}=-6.2$) compared to the model including TiO/VO. This suggests that TiO/VO may  present in the photosphere. However, considering that the temperature inversion extends to the upper layers where TiO/VO are expected to be further dissociated, TiO/VO are not likely to be significantly involved in the temperature inversion.}
We also attempt to fit the data without the ${\rm H^-}$ ion model to evaluate the models with different components in the KELT-9 b's dayside atmosphere. The differences of log marginalized evidence for the model without ${\rm H^-}$ ion is \textcolor{black}{$\Delta {\rm ln}\,{\mathcal Z}=-8.7$} compared to the basic free chemistry model. \textcolor{black}{The presence of the ${\rm H^-}$ ion can be confirmed from the spectrum by the turnoff of the flux at $1.4\ \mu {\rm m}$ and the subsequent decrease in flux at wavelengths longer than $1.4\ \mu {\rm m}$ (see also \cite{Jacob}). This spectral feature is caused by the decrease in the bound-free absorption cross-section of the ${\rm H^-}$ ion at wavelengths longer than $1.4\ \mu {\rm m}$.} This result indicate that ${\rm H^-}$ ions are strongly preferred by the retrievals.
The equilibrium chemistry model has a lower dimension than the free chemistry model, and TiO and VO abundances decrease rapidly with increasing altitude due to thermal dissociation, explaining why it is favored over the free chemistry case. This result is also consistent with the HRCCS result because HRCCS is sensitive to a lower-pressure region. In contrast, the temperature inversion extends up to the TiO/VO-depleted regions.
These results denote that the effect of TiO and VO on the temperature inversion is not likely to be significant in the KELT-9 b dayside atmosphere while the large amount of ${\rm H^-}$ ion may be able to induce the temperature inversion. 
\textcolor{black}{\citet{parmentier} and \citet{Lothringer} classified UHJs by the existence or absence of temperature inversion and optical absorbers in their dayside atmospheres. For planets classified in the highest temperature group ($T_{\rm day}>2700 \rm{K}$), the ${\rm H^-}$ ion acts as the main optical absorber, creating a temperature inversion. This is because, while TiO and VO are thermally dissociated, ${\rm H^-}$ ions are abundantly produced by the thermal dissociation of hydrogen molecules and metal atoms. KELT-9 b belongs to this group, and the low abundances of TiO and VO and the discovery of abundant ${\rm H^-}$ found in this study are consistent with self-consistent models. Moreover, KELT-9 b's host star is an early-type star, and the atomic metals abundantly discovered in studies (\cite{Kasper}) may contribute to the temperature inversion as optical absorbers (\cite{Lothringer-2019}).}
However, there are large residuals between the best-fit model and the observed spectra, especially around $1.4\ \mu {\rm m}$ and $1.6\ \mu {\rm m}$ (figure \ref{fig:residuals}).
Metals and other molecular species not considered in this retrieval may better explain the spectra, while the inclusion of these additional species could potentially affect the abundances of TiO and VO. \textcolor{black}{In addition, the line lists for TiO and VO used in this study have only been calculated up to 3500 K, even though we used the most recent versions available. However, the temperatures calculated in our retrieval reach up to 6000 K. In this case, TauREx3 uses the 3500 K line lists, which may affect the derived abundances of TiO and VO.}

\subsubsection{\textcolor{black}{The abundances of species other than TiO and VO}}
We include the $\rm{H_2O}$, CO, and FeH in the free chemistry model. \textcolor{black}{The retrieved abundances for these molecules are ${\rm log(H_2O)} = -9.16^{+0.71}_{-0.87}$, ${\rm log(CO)} = -5.68^{+0.64}_{-0.57}$, and ${\rm log(FeH)} = -8.48^{+0.30}_{-0.24}$. These measured abundances are almost consistent with the theoretical prediction (e.g., \cite{Kitzmann}) and previous analysis (\cite{changeat2021}; \cite{changeat_edwards2022}).} We confirm the dissociation of $\rm{H_2O}$.  Water is the main outgoing longwave radiation carrier, and the absence of water inhibits cooling by radiation, thereby enhancing the thermal inversion (\cite{Lothringer}). \textcolor{black}{Our retrieval indicates a substantial presence of CO in the atmosphere. The constraint on CO abundance in our model primarily comes from Spitzer's 4.5 $\mu$m photometric data point. The persistence of CO at high temperatures is likely due to its strong triple bond, making it resistant to thermal dissociation.}
FeH, which is thought to cause temperature inversion along with TiO and VO, has such a low abundance that it can be considered a non-detection.

\subsubsection{Metallicity and C/O ratio}
The derived metallicity and C/O ratio are \textcolor{black}{$\rm{[M/H]}=1.04^{+0.20}_{-0.34}$ and $\rm{C/O}=0.37^{+0.26}_{-0.19}$} of the equilibrium retrieval. Metallicity value differs from the  \citet{Jacob} derived value (${\rm [M/H]}=1.98^{+0.19}_{-0.22}$, ${\rm C/O}=0.14^{+0.10}_{-0.03}$) but are consistent in terms of supersolar metallicity. We attempt to reproduce \citet{Jacob} results by excluding the MuSCAT2, Sinistro, and CHEOPS data from the data set but are unable to reproduce \citet{Jacob} results of the equilibrium chemistry model. Differences of values may be due to differences in the retrieval codes used. Derived values are completely different from \citet{Ridden-Harper} derived constraint (${\rm [M/H]}<-0.32$, ${\rm C/O}>0.84$) from HRCCS observation. \textcolor{black}{This discrepancy may be attributed to several factors. Their metallicity constraint is primarily derived from atomic and ionic metal lines, while our metallicity estimate is mainly based on oxygen abundance. Additionally, their analysis showed limited sensitivity to the C/O ratio. Moreover, the HRCCS observations probe lower pressure regions of the atmosphere compared to our low-resolution spectroscopy. At these lower pressures, chemical disequilibrium processes may play a more significant role, potentially leading to different values of metallicity and C/O ratio.}\par
The high metallicity and low C/O ratio derived in this study can be formed through various processes (\cite{madhusudhan14}). \textcolor{black}{Considering the spin-orbit misalignment of KELT-9 b (\cite{spin-orbit}), disk-free migration should be considered. Under such circumstances, possible solutions to achieve high metallicity and low C/O ratio are gravitational instability formation and core erosion after the core accretion formation (\cite{madhusudhan17}).} However, when considering the thermal dissociation of ${\rm H_2O}$ due to the high-temperature environment and the systematics involved in interpreting the combined observational results from various instruments (e.g., \cite{Yip}), the derived metallicity and C/O ratio may be biased values. 
JWST and future telescopes like Ariel could measure the metallicity and C/O ratio more accurately to constrain the formation scenario of KELT-9 b.

\section{Summary}
\label{sec:sum}
We observed secondary eclipses of KELT-9 b using MuSCAT2 in the $g$, $r$, $i$, and $z_{\rm s}$-band and Sinistro in the $Y$-band. \textcolor{black}{We detect secondary eclipses in $i$ and $z_{\rm s}$-bands with more than $3\sigma$ significance. The derived eclipse depths are $373^{+74}_{-75}$ ppm and $638^{+199}_{-178}$ ppm, respectively.}\par
In atmospheric retrievals, we include the data of HST/WFC3, Spitzer, TESS, and CHEOPS. We use free chemistry and equilibrium chemistry models to determine the mixing ratios of TiO and VO. In both approaches, we retrieved a strong temperature inversion, however, we derive less abundance of these molecules than \citet{changeat2021} and \citet{changeat_edwards2022}. This result indicates that temperature inversion in the dayside of KELT-9 b would not originate from TiO and VO, while the large abundance of ${\rm H^-}$ ion may play this role. \textcolor{black}{We also derive the high metallicity and low C/O ratio, which agree with the gravitational instability and core erosion after the core accretion formation scenario.} \par
It should be noted that even the best-fit model does not fully explain the HST/WFC3 spectrum. Molecules and atoms not considered during retrievals may provide a better solution. Increased wavelength coverage \textcolor{black}{as well as higher spectral resolution} would more accurately constrain the abundances of TiO and VO.



\begin{ack}
YH thanks Dr. Yuichi Ito and Dr. Kazumasa Ohno for their helpful advice on the retrieval analysis.
This work is supported by JSPS KAKENHI Grant Numbers JP18H05439, 	20J21872, 21K13955, 21K20376, and JST CREST Grant Number JPMJCR1761. QC was supported by the European Research Council, Horizon 2020 (758892 ExoAI), the STFC/UKSA Grant ST/W00254X/1, and NAOJ Research Coordination Committee, NINS, NAOJ-RCC2301-0401. EP acknowledges funding from the Spanish Ministry of Economics and Competitiveness through project PID2021-125627OB-C32. E. E-B. acknowledges financial support from the European Union and the State Agency of Investigation of the Spanish Ministry of Science and Innovation (MICINN) under the grant PRE2020-093107 of the Pre-Doc Program for the Training of Doctors (FPI-SO) through FSE funds. G. M. acknowledges financial support from the Severo Ochoa grant CEX2021-001131-S and from the Ram\'on y Cajal grant RYC2022-037854-I funded by MCIN/AEI/10.13039/501100011033. JL partly supported by Astrobiology Center SATELLITE Research project AB022006. This article is based on observations made with the MuSCAT2 instrument, developed by ABC, at Telescopio Carlos Sánchez operated on the island of Tenerife by the IAC in the Spanish Observatorio del Teide.

\end{ack}

\appendix
\section*{Additional figures}
\begin{figure*}[H]
 \begin{center}
  \includegraphics[width=16cm]{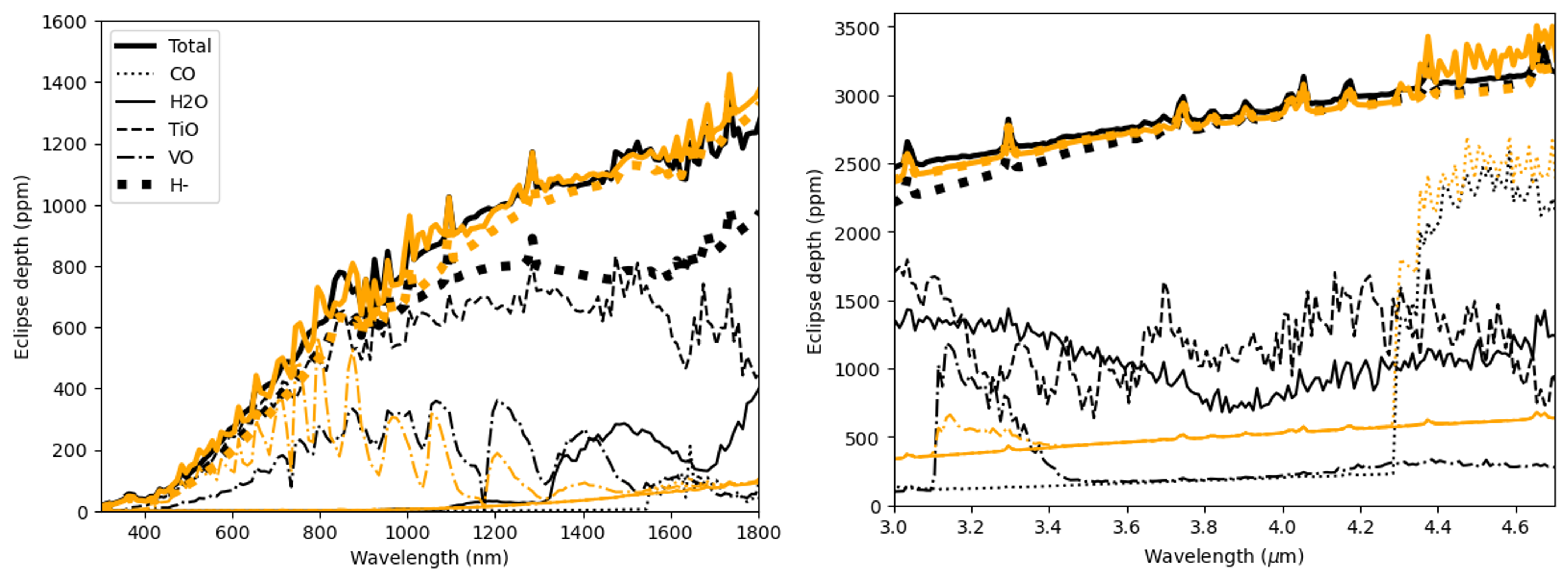}
 \end{center}
 \caption{Contribution of each species. The line type combinations in the figure are the same as in Figure \ref{fig:result}. }\label{figA:contribution}
\end{figure*}

\begin{figure*}[H]
 \begin{center}
  \includegraphics[width=16cm]{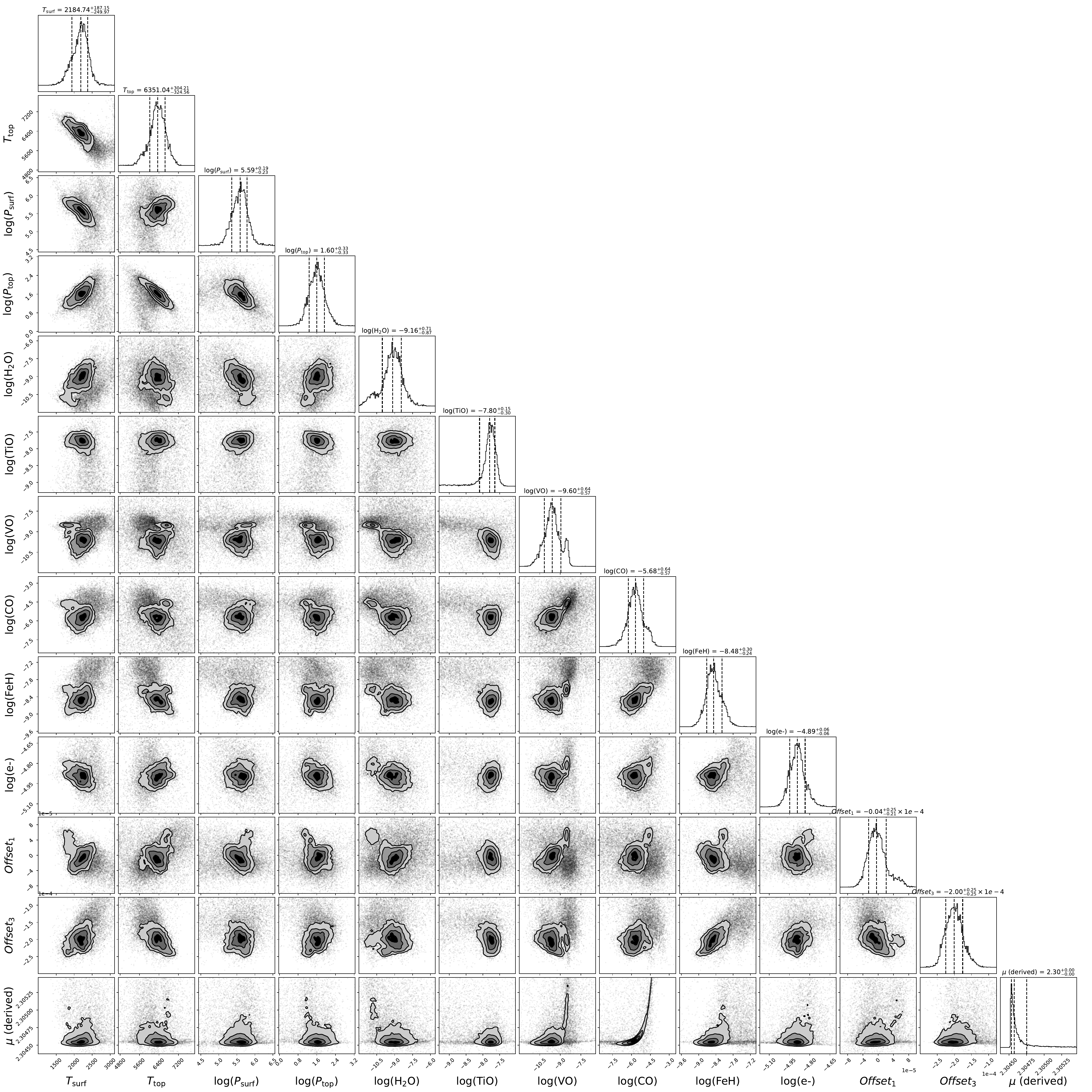}
 \end{center}
 \caption{The posterior distributions of the retrieved atmospheric parameters from the free chemistry using TauREx3. Offset1 and Offset 3 are offsets of MuSCAT2/Sinistro and HST/WFC3, respectively.}\label{figA:FREE_corner}
\end{figure*}

\begin{figure*}[H]
 \begin{center}
  \includegraphics[width=16cm]{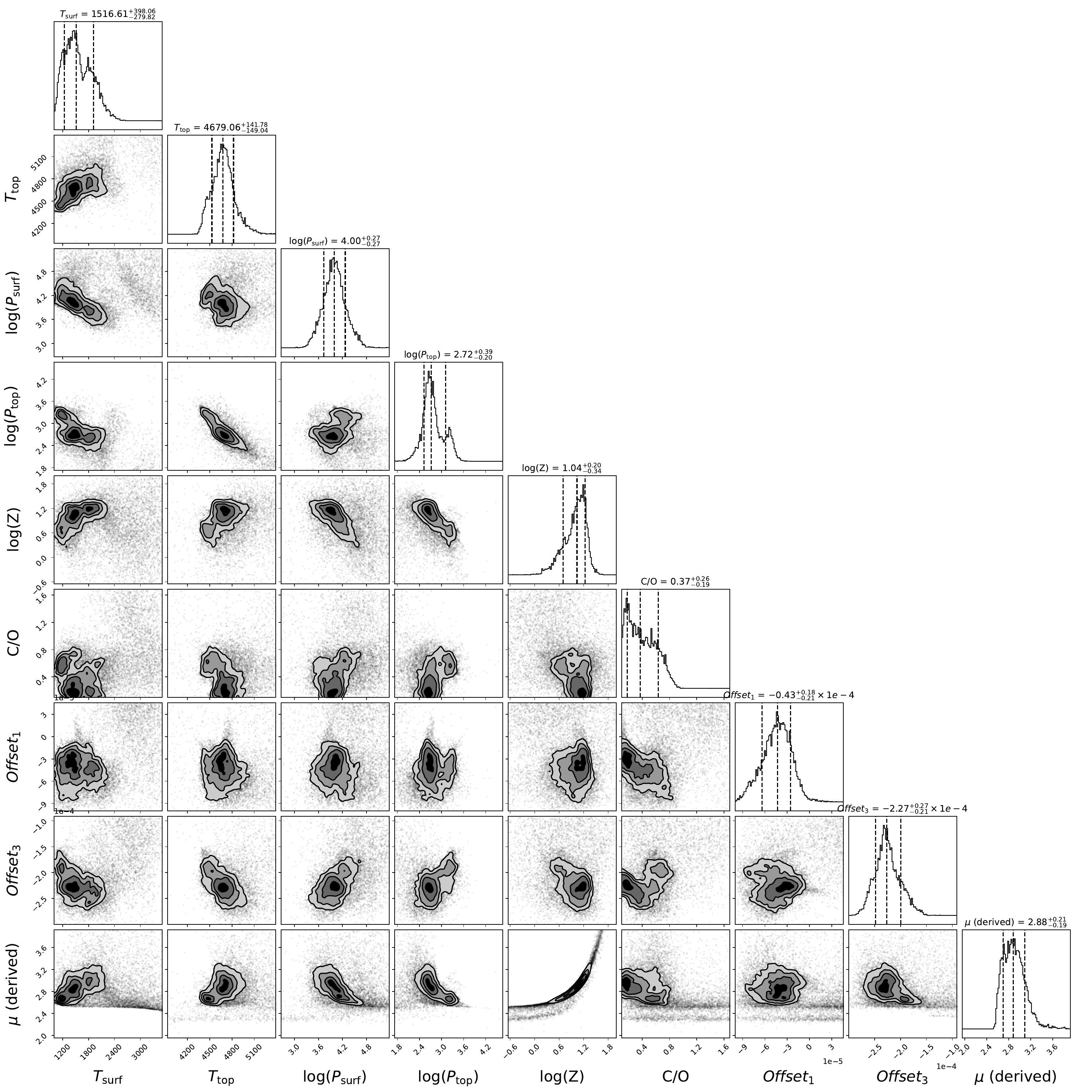}
 \end{center}
 \caption{The posterior distributions of the retrieved atmospheric parameters from the equilibrium chemistry using TauREx3. Offset1 and Offset 3 are offsets of MuSCAT2/Sinistro and HST/WFC3, respectively.}\label{figA:EQ_corner}
\end{figure*}

\textcolor{black}{Contribution of each species are shown in Figure \ref{figA:contribution}.} The posterior distributions of atmospheric retrievals are shown in figure \ref{figA:FREE_corner} and figure \ref{figA:EQ_corner}. Figure \ref{figA:FREE_corner} shows the corner plot of the free chemistry model, while figure \ref{figA:EQ_corner} shows the corner plot of the equilibrium chemistry model.

\bibliography{ref}{}
\bibliographystyle{aasjournal}

\end{document}